\documentclass[12pt]{iopart}
\usepackage{iopams}
\usepackage{setstack}
\usepackage{amssymb}
\usepackage{latexsym}
\usepackage{epsfig}
\usepackage{color}
\usepackage{float}
\usepackage{graphicx}
\usepackage{bm}
\usepackage{verbatim}
\usepackage{cite}

\DeclareMathAlphabet{\mathpzc}{OT1}{pzc}{m}{it}
\makeatletter
\DeclareRobustCommand{\text}{%
  \ifmmode\expandafter\text@\else\expandafter\mbox\fi}
\let\nfss@text\text
\def\text@#1{{\mathchoice
  {\textdef@\displaystyle\f@size{#1}}%
  {\textdef@\textstyle\f@size{#1}}%
  {\textdef@\textstyle\sf@size{#1}}%
  {\textdef@\textstyle \ssf@size{#1}}%
  \check@mathfonts
  }%
}
\def\textdef@#1#2#3{\hbox{{%
                    \everymath{#1}%
                    \let\f@size#2\selectfont
                    #3}}}
\makeatother

\begin{document}

\title{Thermodynamics of Black Holes in Rastall Gravity}
\author{Iarley P. Lobo$^{1}$\footnote{iarley\_lobo@fisica.ufpb.br}, H. Moradpour$^{2}$\footnote{h.moradpour@riaam.ac.ir}, J. P. Morais Gra\c ca$^{1}$\footnote{jpmorais@gmail.com}, I. G. Salako$^{3}$\footnote{inessalako@gmail.com}}
\address{$^{1}$ Departamento de F\'{i}sica, Universidade Federal da Para\'{i}ba, Caixa Postal 5008, CEP 58051-970, Jo\~{a}o Pessoa, PB,
Brazil\\
$^2$ Research Institute for Astronomy and Astrophysics of Maragha
(RIAAM), P.O. Box 55134-441, Maragha, Iran\\
$^{3}$ Institut de Math\'ematiques et de Sciences Physiques (IMSP)
01 BP 613 Porto-Novo, B\'enin}

\begin{abstract}
A promising theory in modifying general relativity by violating the
ordinary energy-momentum conservation law in curved spacetime is
the Rastall theory of gravity. In this theory, geometry and matter fields are
coupled to each other in a non-minimal way. Here, we study
thermodynamic properties of some black hole solutions in this
framework, and compare our results with those of general
relativity. We demonstrate how the presence of these matter sources amplifies effects caused by the Rastall parameter in thermodynamic quantities. Our investigation also shows that black holes with radius
smaller than a certain amount ($\equiv r_0$) have negative heat
capacity in the Rastall framework. In fact, it is a lower bound for
the possible values of horizon radius satisfied by stable black
holes.
\end{abstract}
\maketitle
\section{Introduction}

One of the big puzzles in science is the fact that our universe is going through a phase of accelerated expansion. A possible explanation for this is the presence of a dark energy field, one with constant positive energy density and negative pressure, that can provide a kind of matter that allows such acceleration. Despite the fact
that the nature of the dark sectors of the cosmos is currently unknown, standard cosmology is very successful in describing the cosmos history.
One of these models, based on the presence of a cosmological constant term in Einstein's gravity, is called $\Lambda CDM$, or $\Lambda$
Cold Dark Matter, where $\Lambda$ is the cosmological constant and
plays the role of dark energy. The most promising explanation for
the existence of a cosmological constant is the vacuum energy of
elementary particles, but calculations tell us that such vacuum
energy is more than one hundred orders of magnitude higher than
the measured value for $\Lambda$.

Another possibility is that the negative pressure is generated by
some peculiar kind of perfect fluid, where the proportion between
the pressure and energy density is between $-1$ and $-1/3$. If
this perfect fluid is generated by a scalar field, it is generally
called quintessence, and it is considered as a hypothetical form
of dark energy. Therefore, if this is the case, we must consider
that our universe is pervaded by such fluid, and we must study
strong gravity objects such as black holes in contact with them.
This is the idea presented by Kiselev in \cite{Kiselev:2002dx}, and
has been generalized to the Rastall model of gravity
\cite{rastall} in Ref.~\cite{Heydarzade:2017wxu}.

The idea behind Rastall model is that our laws of conservation,
such as conservation of mass/energy, has been probed only in the
flat or weak-field arena of spacetime \cite{rastall}. A new
generalization of this theory has recently been proposed,
introducing the coupling between matter and gravitational fields
in a non-minimal way as an origin for the accelerating phase of
the universe \cite{genras}. Based on Rastall's argument, the necessity
that the covariant derivative of the energy-momentum tensor to be
zero can be relaxed, allowing one to add new terms to the
Einstein's equation. In fact, it has recently been shown
that the divergence of the energy-momentum tensor
can be non-zero in a curved spacetime \cite{PRL}. For this theory,
several exact solutions has been obtained, both for astrophysical
\cite{REPJC,deMello:2014hra,Oliveira:2015lka,Bronnikov:2016odv,vel,Heydarzade:2016zof,Licata,Heydarzade:2017wxu,gaus}
and cosmological scenarios
\cite{Capone:2009xm,Batista:2011nu,Silva:2012gn,Santos:2014ewa,hm,Yuan:2016pkz,RHAG}.

Comparing thermodynamic quantities and properties of black holes
in Rastall gravity with their counterparts in general relativity
helps us to be more familiar with the nature of a non-minimal
coupling between geometry and matter fields introduced in the
Rastall hypothesis. Besides, thermodynamics properties of Kiselev
solutions in the general relativity framework have been studied by
some authors \cite{p0,p1,p2,p3,p4,p5}. In fact, thermodynamic
properties of diverse black holes have extensively been studied in
various theories of gravity
\cite{1,2,3,4,5,6,7,8,9,10,11,12,13,14,15,16,17,18,19,20,21,22,23,24,25}.
Hence, our aim in this paper is to study the thermodynamic
quantities and properties of black holes surrounded by a
prefect fluid in the Rastall framework. An appealing property that we found concerns the coupling of the Rastall parameter with the energy density of the surrounding fields: they couple in such a way that the fields densities work as an amplifier to Rastall-like deformations, which could help us to experimentally distinguish between this formalism and general relativity.

This paper is organized as follows. In the next section, we
address general remarks on black hole solutions surrounded by a
perfect fluid in general relativity and the Rastall theory as well
as the thermodynamic quantities of black holes in the Rastall
framework. In sections \ref{seciii} and \ref{seciv}, energy and pressure of
solutions surrounded by a quintessence field and cosmological
constant, as promising approaches to describe the current
accelerating universe, respectively, are studied in details in the Einstein and Rastall frameworks. The case of phantom
field is also investigated in section \ref{secv}. In section
\ref{secvi}, we will study the possibility of the occurrence of
phase transitions in Rastall black holes. The last section is
devoted to a summary and concluding remarks.


\section{A black hole surrounded by a perfect fluid in Rastall gravity; general remarks}

In this paper we will work with a Schwarzschild-like metric, given by

\begin{equation}
ds^2 = - f(r) dt^2 + \frac{1}{f(r)} dr^2 + r^2 d \Omega^2,
\label{metric}
\end{equation}

\noindent with $d\Omega^2  = d\theta^2 + \sin^2(\theta)
d\phi^2$, along with a general spherically symmetric energy-momentum tensor. The general expression for time and spatial components
of such tensor is given by

\begin{equation}
T^t_t= A(r),\  T^t_i = 0,\ and\ T^i_j= C(r)r_j r^i + B(r)
\delta^i_j.
\end{equation}

For general relativity, Kiselev has shown \cite{Kiselev:2002dx} that if the background is
filled by a source with pressure $p(r)$ and energy density
$\rho(r)$, related to each other by a state parameter
$\omega_q\equiv\frac{p(r)}{\rho(r)}$, then

\begin{equation}
f(r) = 1 - \frac{2M_K}{r} - N_K r^{\eta_K}
\end{equation}

\noindent in which $M_K$ and $N_K$ are constants of integration
and

\begin{equation}
\eta_K = -1-3\omega_q.
\end{equation}


\subsection{Black holes surrounded by a perfect fluid in Rastall
gravity}

Rastall gravity is a theory where the total energy-momentum
tensor is not conserved, but its covariant derivative is
proportional to the derivative of the Ricci scalar \cite{rastall}. This means that,
in a flat spacetime or as a first approximation of a weak gravitational field, all the known laws of conservation are valid.
We should stress that such laws has been tested only in the weak
regime of gravity, and it is known that gravity can produce
particles via quantum effects, thus breaking some of these laws
\cite{rastall}.

Rastall hypothesis can be written as

\begin{equation}
\nabla_\mu T^{\mu\nu} = \lambda \nabla^\nu R, \label{Rastall1}
\end{equation}

\noindent where $\lambda$ is the Rastall parameter, and general relativity is recovered in the limit $\lambda \rightarrow 0$. From
(\ref{Rastall1}), we can write the modified equations of gravity
as

\begin{equation}
H^\mu_\nu \equiv G^\mu_\nu + \lambda \kappa \delta^\mu_\nu R =
\kappa T^\mu_\nu, \label{Rastall2}
\end{equation}

\noindent where $\kappa$ is Rastall's gravitational constant. To
find solutions to these field equations, one should
solve the set of equations (\ref{Rastall2}) for some
energy-momentum tensor.

Taking the trace of equation (\ref{Rastall2}), we have

\begin{equation}
R (4\lambda\kappa - 1) = \kappa T,
\end{equation}

\noindent which means that in vacuum one must have $R=0$ or $\kappa\lambda=1/4$. As the latter option is not allowed (see
\cite{rastall,Moradpour:2016fur}), the former should be the case, and we
must have that all vacuum solutions in GR are also solutions for Rastall gravity.

Applying Kiselev's approach \cite{Kiselev:2002dx} to the
field equations, and considering $\rho(r) = A r^{\beta}$,
where both $A$ and $\beta$ are constants, Heydarzade and Darabi
\cite{Heydarzade:2017wxu} found out

\begin{equation}
\beta = -
\frac{3(1+\omega_q)-12\kappa\lambda(1+\omega_q)}{1-3\kappa\lambda(1+\omega_q)},
\end{equation}

\noindent and

\begin{equation}
A = \frac{3N (1- 4 \kappa \lambda)(\kappa
\lambda(1+\omega_q)-\omega_q)}{\kappa(1-3\kappa\lambda(1+\omega_q))^2},
\label{Aconstant}
\end{equation}

\noindent where $N$ is an integration constant, related with the
surrounding field. Hence, the metric function is given by

\begin{equation}
f(r) = 1 - \frac{2M}{r} - N r^\eta, \label{ffunction}
\end{equation}

\noindent with

\begin{equation}\label{eta0}
\eta = -
\frac{1+3\omega_q-6\kappa\lambda(1+\omega_q)}{1-3\kappa\lambda(1+\omega_q)},
\end{equation}

\noindent where $M$ is another integration constant, representing
the black hole mass. For $\lambda \rightarrow 0$ we recover the
solution found by Kiselev \cite{Kiselev:2002dx} in the framework
of general relativity. It is worthwhile mentioning that similar
solutions can also be obtained in Rastall framework for other
situations \cite{vel,Heydarzade:2016zof}.

For each choice of equation of state, we can find the metric, as
given by (\ref{ffunction}), along with the constant $A$, given by
(\ref{Aconstant}). To preserve the weak energy condition, i.e.,
$\rho > 0$, we must have $A > 0$, and this will define the sign of
the parameter $N$ as dependent on the parameter $\kappa \lambda$.
For a full discussion on each case, see \cite{Heydarzade:2017wxu}.


\subsection{Thermodynamic quantities of black holes in Rastall Gravity}

To work on the thermodynamic aspects of the Rastall model of
gravity, we must define classical thermodynamic quantities, such
as energy and entropy. These are local quantities, but for general
relativity there is no straightforward way (it is even senseless)
to define the local energy of a gravitational field configuration.
For the total energy, the most accepted ones are the ADM energy at
spatial infinity \cite{Arnowitt:1959ah}, and the Bondi-Sachs
\cite{Bondi:1962px,Sachs:1962wk} energy at null infinity, both
describing an isolated system in an asymptotically flat spacetime.
But its local counterpart should be chosen as a useful quantity,
defined for the interior of some well-defined boundary, that goes
to one of the well-accepted values of energy as the boundary goes
to infinity.

Some useful definitions for this quasi-local notion of energy
exist in literature \cite{revm}, and in this paper we will use the results
obtained in \cite{Moradpour:2016fur}, that uses a generalized
Misner-Sharp definition of energy \cite{Misner:1964je}, as a suitable definition for energy \cite{j1,j2,j3,j4,j5,j6,j7,j8,j9,j10,j11,j12,j13} to find
the corresponding entropy related with the geometry of the
spacetime in Rastall gravity. The unified first law of thermodynamics (UFL) is defined as \cite{ref2}

\begin{eqnarray}\label{ufl}
dE\equiv A\Psi_a dx^a + W dV,
\end{eqnarray}

\noindent in which, $\Psi_a\, =\, T^b_a\partial_b r + W\partial_a r$ is the energy supply vector, and $W=-\frac{h^{ab}T_{ab}}{2}$ denotes the work density. Moreover, $h_{ab}=\text{diag}(-f(r),\frac{1}{f(r)})$ for metric~(\ref{metric}), and in fact, it is the metric on two dimensional hypersurface $(t,r)$. UFL is compatible with the generalization of the Misner-Sharp mass \cite{ref2}, and therefore, one can use the above equation in order to find the generalized Misner-Sharp mass in the gravitational theory under investigation \cite{ref2,Moradpour:2016fur}. Defining $\gamma \equiv \lambda \kappa$, the Newtonian limit leads to \cite{Moradpour:2016fur}

\begin{equation}\label{kap}
\kappa = \frac{4 \gamma-1}{6\gamma -1} 8\pi\ ,
\end{equation}

\noindent and applying the Unified first law of thermodynamics to
the horizon of metric~(\ref{metric}), one can get the Misner-Sharp
mass content of black holes in Rastall gravity as \footnote{For a general derivation of thermodynamic quantities in Rastall gravity, see
\cite{Moradpour:2016fur}.}

\begin{equation}
E = \frac{6\gamma-1}{2(4\gamma-1)}[(1-2\gamma)r_H + \gamma r_H^2
f'(r_H)], \label{energy}
\end{equation}

\noindent where $'$ means derivative with respect to the
coordinate $r$, and $r_H$ denotes the horizon radius. In order to obtain the system pressure, one can use the $r-r$ component of the Rastall field equations~(\ref{Rastall2}) to obtain \cite{Moradpour:2016fur}

\begin{eqnarray}\label{pressure0}
&P(r_H) = \frac{6\gamma-1}{(4\gamma-1) 8\pi}\left(\frac{1}{r_H}[r_H
f'(r_H)-1]-\frac{\gamma}{r_H^2}[r_H^2 f''(r_H) + 4r_H
f'(r_H)-2]\right).
\end{eqnarray}

\noindent In addition, bearing the first law of thermodynamics ($dE=TdS-PdV$) in mind, and using Eqs.~(\ref{energy}) and~(\ref{pressure0}), it has been shown that the entropy of black hole is \cite{Moradpour:2016fur}

\begin{equation}\label{ent}
S = \left(1 + \frac{2\gamma}{4\gamma-1}\right)S_o.
\end{equation}

\noindent Here, $S_o = A/4$ is the well-known
Bekenstein entropy. We can note that, as $\gamma \rightarrow 0$,
one recovers the formulas valid in general relativity
\cite{Moradpour:2016fur}. It is useful to note here that since the $\gamma=\frac{1}{4}$ case is not allowed in this theory \cite{rastall,Moradpour:2016fur}, the singularity of the above relations at this value of $\gamma$ is not worrying.


\section{Thermodynamics of a black hole surrounded by a quintessence field in Rastall gravity}\label{seciii}
\par
A quintessence field, with state parameter $\omega_q$ ranging
between $-1<\omega_q<-1/3$, may be responsible for the observed
accelerated expansion of the universe
\cite{Kiselev:2002dx,Vikman:2004dc} (for a review, see
\cite{Sahni:2004ai}). Here, we will consider the case of a black
hole surrounded by the quintessence field with $\omega_q=-2/3$,
such that
\begin{equation}
f(r)=1-\frac{2M}{r}-N_q r^{\frac{1+2\gamma}{1-\gamma}},
\end{equation}

\noindent which reproduces the solution of Ref.
\cite{Kiselev:2002dx} for $\gamma=0$. Applying this state
parameter to Eq. (\ref{energy}), the Misner-Sharp mass content
confined in the horizon is

\begin{equation}\label{energy1}
E_q=\frac{(1-6\gamma)}{8}\left[\frac{r_H(1-\gamma)^2-N_q\,
\gamma(2+\gamma)r_H^{\frac{2+\gamma}{1-\gamma}}}{(\gamma-1)(\gamma-\frac{1}{4})}\right],
\end{equation}

\noindent which is equal to the Schwarzschild case $E=r_H/2$, for
$\gamma=N_q=0$.
\par
\begin{figure}[H]
\includegraphics[scale=0.45]{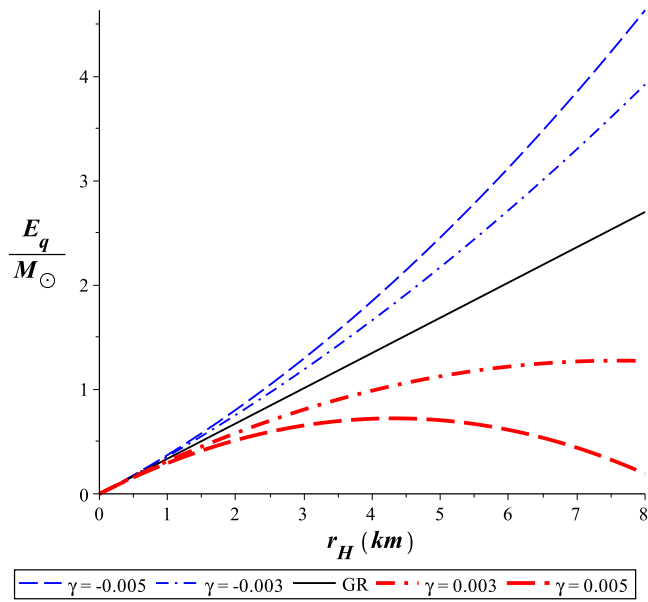}
\caption{Misner-Sharp mass for quintessence field $E_q$ as a
function of the horizon $r_H$, in solar mass units. We considered
the cases of $\gamma=-0.005$ (blue, dashed line), $\gamma=-0.003$
(blue, dash-dotted line), GR (black, solid line), $\gamma=0.003$
(red, dash-dotted, thick line) and $\gamma=0.005$ (red, dashed,
thick line). We used $N_q=0.01$ and $M_{\odot}\, G /c^2\approx
1.48\, \text{km}$ is the solar mass in length units.}
\label{energyXrh1}
\end{figure}

We plotted $E_q$ as a function of the horizon radius ($r_H$) for
different values of the Rastall parameter $\gamma$ in
Fig.~(\ref{energyXrh1}). One can see that for small positive values
of $\gamma$, the energy grows until a maximum value is reached, then
diminishes indefinitely and eventually becomes negative from a
finite value of $r_H$. Such behavior is absent in general
relativity, where a linear energy growth is observed. The coupling
between the Rastall parameter $\gamma$ and the quintessence energy
density parameter $N_q$ is essential for this kind of behavior.
\par
Instead, for negative small values of $\gamma$, Rastall gravity
furnishes an ever positive contribution to the Misner-Sharp mass. Considering deviations of GR like
$\gamma=\pm 0.003$, $\gamma=\pm 0.005$ and $N_q=0.01$, we plotted it
just up to $r_H\sim 10\, \text{km}$, because this is the order of
magnitude of the detected BHs by LIGO
\cite{Abbott:2016blz,Abbott:2016nmj,Abbott:2017vtc,Abbott:2017oio}.
In these cases, the energy content within the horizon is increased
in comparison to GR for a fixed value of the horizon. The energy
density of the quintessence fluid is proportional to $N_q$ (as can
be seen in \cite{Heydarzade:2017wxu}) and for GR the formula for
the Misner-Sharp mass does not depends on $N_q$, therefore the
higher the quintessence energy density, the higher is the
difference between GR and Rastall gravity.

The pressure at the horizon is found from (\ref{pressure0}) as
\begin{eqnarray}
P_q=-\frac{N_q}{8\pi}\frac{(2+\gamma)(1-6\gamma)}{(1-\gamma)^2}r_H^{\frac{4\gamma-1}{1-\gamma}}.
\end{eqnarray}

\begin{figure}[H]
\includegraphics[scale=0.55]{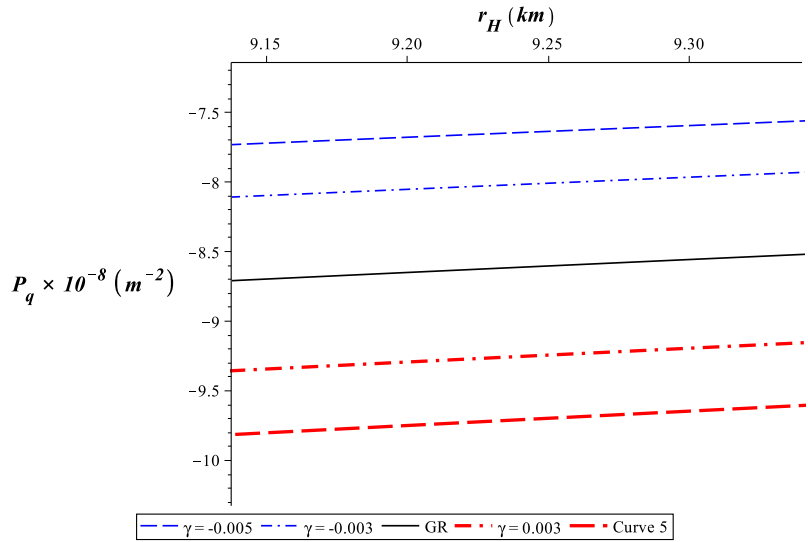}
\caption{The pressure for quintessence field $P_q$ as a function
of the horizon $r_H$. We considered the cases of $\gamma=-0.005$
(blue, dashed line), $\gamma=-0.003$ (blue, dash-dotted line), GR
(black, solid line), $\gamma=0.003$ (red, dash-dotted, thick line)
and $\gamma=0.005$ (red, dashed, thick line). We used $N_q=0.01$.}
\label{pressureXrh1}
\end{figure}

The quintessence fluid is characterized by presenting a negative
pressure (for a positive energy density). This way, it is expected
that its presence might be described by the negativity of the
system's pressure. This is a feature presented both in GR and
Rastall gravity, however its dependence with the horizon $r_H$ is
affected by the Rastall parameter. For the cases that we are
considering, its qualitative behavior is preserved (as can be seen
in Fig.~(\ref{pressureXrh1})), i.e., the pressure gets suppressed by
the BH horizon's size.
\par
It should be noted that if the Rastall parameter is such that
$(2+\gamma)(1-6\gamma)<0$, its coupling with $N_q$ inverts the
sign of the pressure to be positive. The pressure is negative for
$-2\, <\, \gamma\, <\, 1/6$ and is positive for $\gamma\, <\, -2\,
\cup\, \gamma\, >\, 1/6$. And also the modulus of the pressure
grows with the horizon for $1/4\, <\, \gamma\, <\, 1$ and
decreases for $\gamma\, <\, 1/4\, \cup\, \gamma\, >\, 1$ (which is
the present case).

\section{Thermodynamics of a black hole surrounded by a cosmological constant in Rastall gravity}\label{seciv}

Consider a cosmological constant surrounding the BH, i.e.,
$\omega_c=-1$, that may also presumably drive the accelerated
expansion of the universe \cite{Perlmutter:1998np}. The metric
function is the same for GR and Rastall gravity (it does not depend
on $\gamma$), and is given by

\begin{equation}
f(r)=1-\frac{2M}{r}-N_c\, r^2.
\end{equation}

\noindent In fact, it is a solution obtained from other
considerations \cite{vel,Heydarzade:2016zof}. The Misner-Sharp
mass confined in the horizon is

\begin{equation}\label{energycosmconst}
E_c=\frac{1-6\gamma}{2-8\gamma}\left(1-\gamma-3\gamma N_c\, r_H^2\right)r_H.
\end{equation}

\begin{figure}[H]
\includegraphics[scale=0.47]{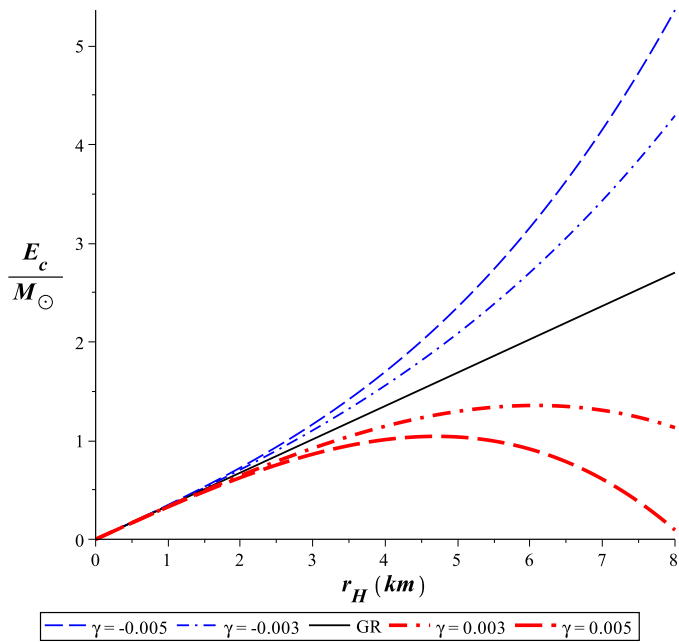}
\caption{Misner-Sharp mass for quintessence field $E_q$ as a
function of the horizon $r_H$, in solar mass units. We considered
the cases of $\gamma=-0.005$ (blue, dashed line), $\gamma=-0.003$
(blue, dash-dotted line), GR (black, solid line), $\gamma=0.003$
(red, dash-dotted, thick line) and $\gamma=0.005$ (red, dashed,
thick line). We used $N_q=10^{-6}$ and $M_{\odot}\, G /c^2\approx
1.48\, \text{km}$ is the solar mass in length units.}
\label{energyXrh2}
\end{figure}

\noindent As can be seen form Eq.(\ref{energycosmconst}), even
thought the metric does not depend on $\gamma$, due to the
modified field equations of this theory, the Misner-Sharp mass is
deformed. The preservation of the metric function is responsible
for avoiding a possible $\gamma$-dependence in the power of $r_H$.
However, as in the previous case, an important extra contribution
arises from the coupling between the energy density $N_c$ and
$\gamma$. For the same reasons of the previous section, and for
the same set of parameters, we depict $E_c$ as a function of $r_H$
in Fig.~(\ref{energyXrh2}). As can be seen, a similar qualitative
behavior is found independently on the fluid under consideration.
From Eq.(\ref{energycosmconst}), for $N_c>0$, the Misner-Sharp
mass is a concave function of $r_H$ for $0\, <\, \gamma\, <\,
1/6\, \cup\, \gamma\, >\, 1/4$, and is a convex one for $\gamma\,
<\, 0\, \cup\, 1/6\, <\, \gamma\, <\, 1/4$.

The pressure at the horizon is a constant

\begin{equation}
P_c=-\frac{3}{8\pi}(1-6\gamma)N_c.
\end{equation}

\noindent It becomes positive for $\gamma\, >\, 1/6$, whenever
$N_C>0$. It is also apparent that a negative pressure is obtainable
for $\gamma\, >\, 1/6$ if $N_C<0$.


\section{Thermodynamics of a black hole surrounded by a phantom field in Rastall gravity}\label{secv}

Another interesting fluid that we analyze consists in the so
called phantom field \cite{Vikman:2004dc,Caldwell:1999ew} with a
super-negative equation of state $\omega_p <-1$. For our purposes,
we consider $\omega_p=-4/3$. Thus the metric function reads
\begin{equation}
f(r)=1-\frac{2M}{r}-N_pr^{\frac{3-2\gamma}{1+\gamma}}.
\end{equation}

The Misner-Sharp mass confined in the horizon is
\begin{equation}
E_p=\frac{(6\gamma-1)}{8}\left[\frac{r_H(1-\gamma^2)+N_p\gamma(\gamma-4)r_H^{\frac{4-\gamma}{1+\gamma}}}{(\gamma+1)(\gamma-\frac{1}{4})}\right].
\end{equation}

\begin{figure}[H]
\includegraphics[scale=0.47]{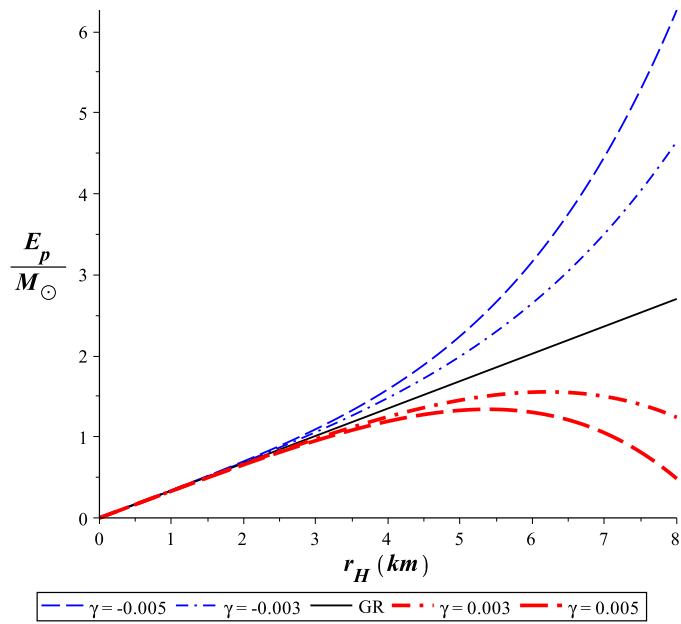}
\caption{Misner-Sharp mass for phantom field $E_p$ as a function
of the horizon $r_H$, in solar mass units. We considered
the cases of $\gamma=-0.005$ (blue, dashed line), $\gamma=-0.003$
(blue, dash-dotted line), GR (black, solid line), $\gamma=0.003$ (red, dash-dotted, thick line) and $\gamma=0.005$ (red, dashed, thick line). We used
$N_p=10^{-10}$.} \label{energyXrh3}
\end{figure}

Qualitatively, it behaves similarly to the quintessence case,
however with a mass growth governed by an approximately fourth
power law, instead of the approximately squared one of the
quintessence field, as is depicted in Fig.~(\ref{energyXrh3}).

The pressure at the horizon is
\begin{eqnarray}
P_p=\frac{-N_p}{8\pi}\frac{(4-\gamma)(1-6\gamma)}{(1+\gamma)^2}r_H^{\frac{1-4\gamma}{1+\gamma}}.
\end{eqnarray}

\begin{figure}[H]
\includegraphics[scale=0.5]{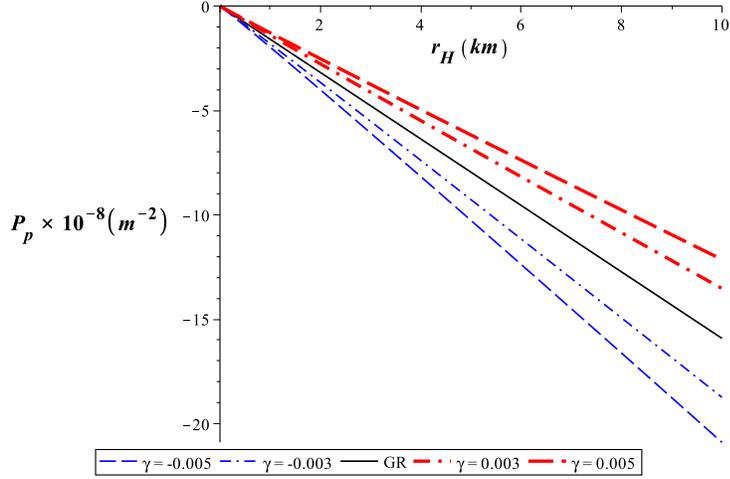}
\caption{The pressure for phantom field $P_q$ as a function of the
horizon $r_H$. We considered the cases of $\gamma=-0.005$ (blue,
dashed line), $\gamma=-0.003$ (blue, dash-dotted line), GR (black,
solid line), $\gamma=0.003$ (red, dash-dotted, thick line) and
$\gamma=0.005$ (red, dashed, thick line). We used $N_p=10^{-10}$.}
\label{pressureXrh2}
\end{figure}

It becomes more negative with the growth of $r_H$. Also, the
distinction between the various values of the Rastall parameter
$\gamma$ becomes more explicit with the increasing of the horizon, as can be seen in Fig.~(\ref{pressureXrh2}).
\par
The pressure is positive for $1/6\, <\, \gamma\, <\, 4$ and is
negative for $\gamma\, <\, 1/6\, \cup\, \gamma\, >\, 4$. Also,
the modulus of the pressure decreases with the horizon for
$\gamma\, <\, -1\, \cup\, \gamma\, >\, 1/4$ and increases for
$-1\, <\, \gamma\, <\, 1/4$ (which is the present case).

\section{Phase transition in black holes}\label{secvi}

The possibility of occurrence of phase transition \cite{path} for
black holes has been studied in various theories of gravity to get
more information on the thermodynamic features of black holes
\cite{p0,p1,p2,p3,p4,p5,1,2,3,4,5,6,7,8,9,10,11,12,13,14,15,16,17,18,19,20,21,22,23,24,25}.
Here, we are going to study the phase transitions for black holes in
Rastall gravity by focusing on Kiselev counterpart solutions
\cite{Kiselev:2002dx} in Rastall gravity
\cite{Heydarzade:2017wxu}. Indeed, these solutions are generally
more than a generalization of the Kiselev solutions to the Rastall
framework, and can also be valid in some other situations
\cite{vel,Heydarzade:2016zof}.

For the spherically symmetric static metric~(\ref{metric}), the
radius of event horizon ($r_H$) can be found by solving the
$f(r_H)=0$ equation. Bearing Eq.~(\ref{ffunction}) in mind, one
can easily see that for $N>0$ and $N<0$, the $\eta=2$ case
recovers the de-Sitter (dS) and anti de-Sitter (AdS) universes,
respectively. Moreover, the Reissner-Nordstr\"om (RN) universe can
be obtained by taking into account the $\eta=-2$ case
\cite{vel,Heydarzade:2016zof}. In fact, if a black hole is
surrounded by a radiation source ($\omega_q=\frac{1}{3}$), then
independently of the value of the parameter $\gamma$, we have $\eta=-2$
\cite{vel,Heydarzade:2016zof}.

Now, since $S = \left(1 + \frac{2\gamma}{4\gamma-1}\right)\pi
r_H^2$, we have $r_H=\alpha\sqrt{S}$, where
$\alpha=\sqrt{\frac{4\gamma-1}{\pi(6\gamma-1)}}$, combined with
Eq.~(\ref{ffunction}) to get

\begin{equation}\label{f2}
f^\prime(r_H)\rightarrow f^\prime(S) = \frac{M^\prime}{S} -
N^\prime S^{(\eta-1)/2}.
\end{equation}

\noindent in which $N^\prime\equiv\eta\alpha^{\eta-1}N$ and
$M^\prime\equiv\frac{2M}{\alpha^2}$. Since the Misner-Sharp
definition of energy is fully compatible with the unified first law of
thermodynamics \cite{Moradpour:2016fur}, we take it as the total
thermodynamic energy of system. Therefore, using
Eq.~(\ref{energy}) and defining
$\alpha^\prime\equiv\frac{(1-2\gamma)}{2\pi\alpha}=(1-2\gamma)\sqrt{\frac{6\gamma-1}{4\pi(4\gamma-1)}}$,
we easily reach at

\begin{equation}\label{e1}
E(S)=\alpha^\prime\sqrt{S} + \frac{\gamma}{2\pi}(M^\prime-N^\prime
S^\frac{\eta+1}{2}).
\end{equation}

The Hawking temperature and the heat capacity can be found as

\begin{equation}\label{T}
T(S)=\frac{dE}{dS}=\frac{\tilde{\alpha}-\tilde{N}S^\frac{\eta}{2}}{\sqrt{S}},
\end{equation}

\noindent and

\begin{equation}\label{C}
C=T\frac{dS}{dT}=\frac{2S(\tilde{\alpha}-
\tilde{N}S^\frac{\eta}{2})}{(\tilde{\alpha}-
\tilde{N}S^\frac{\eta}{2})(\eta-1)-\eta\tilde{\alpha}},
\end{equation}

\noindent where $\tilde{\alpha}=\frac{\alpha^\prime}{2}$ and
$\tilde{N}=\frac{\gamma N^\prime(\eta+1)}{4\pi}$, respectively. Hence, heat capacity diverges at
$S_0=\frac{S_m}{(1-\eta)^\frac{2}{\eta}}$, where
$S_m\equiv(\frac{\tilde{\alpha}}{\tilde{N}})^\frac{2}{\eta}$,
meaning that there can be a second order phase transition at this
point \cite{path}. In fact, by bearing Eq.~(\ref{eta0}) in mind,
one can see that the value of $\eta$, and therefore, the
possibility of occurrence of a phase transition depends on the values
of $\gamma$ and $\omega_q$. As a check, we can easily see that the
results of considering the Schwarzschild metric can be obtained by applying both the
$\gamma\rightarrow0$ and $\eta\rightarrow0$ limits to the above
results.

\begin{figure}[H]
\centering
\includegraphics[width=2.7in, height=2.7in]{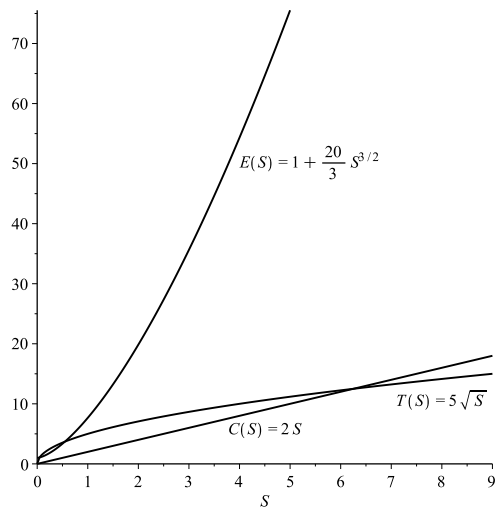}
\caption{\label{fig1} Temperature and heat capacity for $\eta=2$,
$\tilde{N}=-5$ and $\tilde{\alpha}=0$. For the energy curve,
$M^\prime=4\pi$ and $\gamma=\frac{1}{2}$ compatible with
$\tilde{\alpha}=0$ leading to $\omega_q=-1$.}
\end{figure}

It is interesting to note here that if $\tilde{\alpha}=0$ and
$\eta>\frac{1}{2}$, then $T\rightarrow0$ for $S\rightarrow0$. This means
that the second law of thermodynamics is satisfied by this
case. Moreover, $\tilde{N}$ should be negative to meet the $S>0$
condition, meaning that this case requires $\tilde{N}\leq0$.
Besides, heat capacity is positive only if $\eta>1$. Additionally,
since $S_0=0$, there is only one phase with
$E=\frac{\gamma}{2\pi}M^\prime-\frac{2\tilde{N}}{\eta+1}S^\frac{\eta+1}{2}$.
This way, if $\eta=2$ and $\gamma=\frac{1}{2}$, we have
$\alpha^\prime=\tilde{\alpha}=0$ and $\omega_q=-1$. Moreover,
Eq.~(\ref{ent}) implies $S=2S_0$ in this situation. To clarify the
behavior of this case, energy, temperature and heat capacity have
been plotted in Fig.~(\ref{fig1}).

\begin{figure}[H]
\centering
\includegraphics[width=2.7in, height=2.7in]{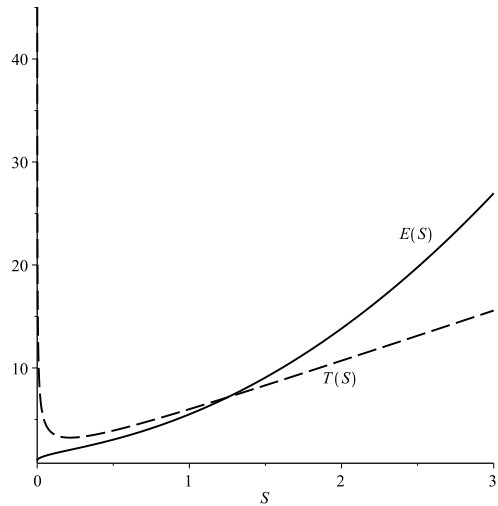}
\caption{\label{fig2} Temperature and Energy for
$M^\prime=\frac{2\pi}{\gamma}$, $\tilde{N}=-5$ and
$\tilde{\alpha}=1$ while $\eta=3$.}
\end{figure}

\begin{figure}[H]
\centering
\includegraphics[width=2.7in, height=2.7in]{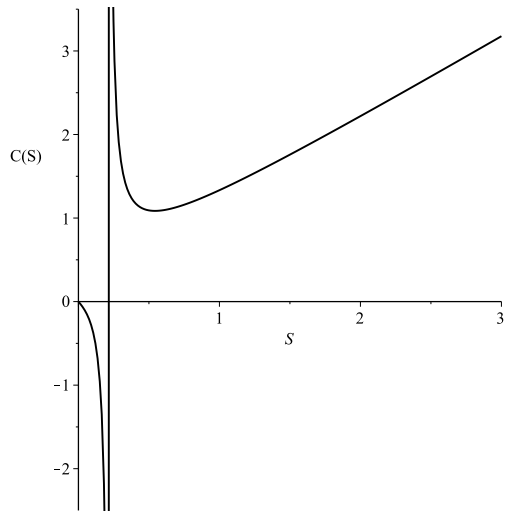}
\caption{\label{fig3} Heat capacity for $\tilde{N}=-5$,
$\tilde{\alpha}=1$ and $\eta=3$. There is a second order phase
transition located at $S_0=(\frac{1}{100})^{\frac{1}{3}}$}
\end{figure}

Now, for the $\frac{\tilde{\alpha}}{\tilde{N}}<0$ case parallel to
$\tilde{\alpha}>0$, since $\tilde{N}$ is negative, the temperature is
positive everywhere, and $T(S\rightarrow0)\rightarrow\infty$,
meaning that the second law of thermodynamics is not satisfied.
While for $\eta<1$, the temperature drops to zero for $S\gg1$. If we have
$\eta>1$, then the temperature have a minimum located at $S=S_0$ for
that $T(S_0)=\frac{\eta\tilde{\alpha}}{(\eta-1)\sqrt{S_0}}$, and
it increases as a function of $S$ for $S>S_0$. It is worthwhile
mentioning that, for $\eta=1$, there is no singularity in the
behavior of heat capacity, and $T(S\gg1)\approx-\tilde{N}$.
Temperature, energy and heat capacity have been plotted in
Figs.~(\ref{fig2}) and~(\ref{fig3}), respectively. Here, we only
focus on the $M^\prime=\frac{2\pi}{\gamma}$ case combined with the
definitions of $M^{\prime}$, $\alpha$, $\gamma$ and
Eq.~(\ref{kap}) to reach at $M=\frac{\lambda}{8\pi G}$.


In Fig.~(\ref{fig2}), we show that while the temperature is
positive for $S<S_0$, its changes are very expressive. Besides, as it is
apparent from Fig.~(\ref{fig3}), the heat capacity is negative for
$S<S_0$, meaning that it is an unstable phase \cite{25}. Therefore,
black holes of radius $r_H<r_0\equiv\alpha\sqrt{S_0}$ are unstable
and, in fact, $r_0$ is a lower bound for the radius of a stable black
holes in this approach.

\begin{figure}[H]
\centering
\includegraphics[width=2.7in, height=2.7in]{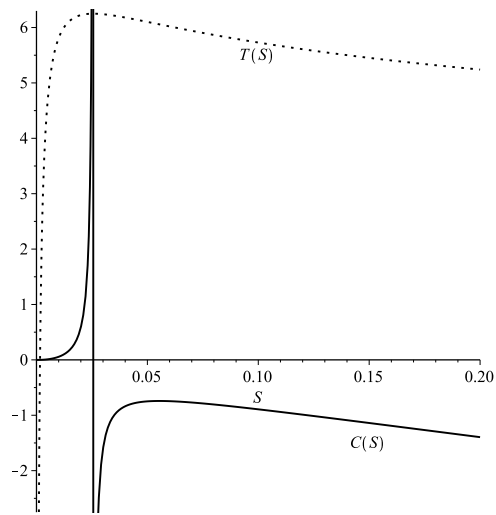}
\caption{\label{fig4} Temperature and heat capacity for
$\tilde{N}=-5$ and $\tilde{\alpha}=-1$ while $\eta=\frac{1}{2}$
and thus $S_m=(\frac{1}{5})^{4}$.}
\end{figure}

\begin{figure}[H]
\centering
\includegraphics[width=2.7in, height=2.7in]{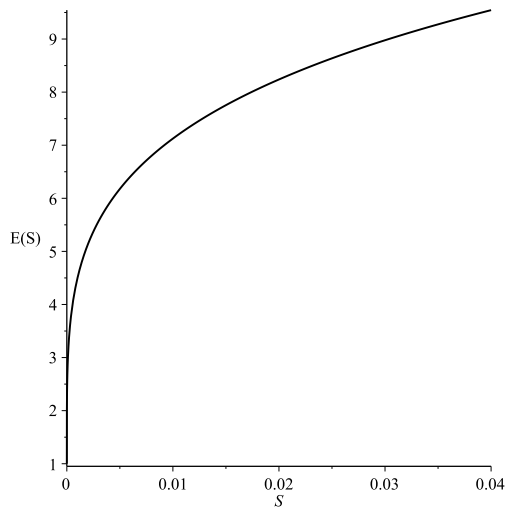}
\caption{\label{fig5} Energy for $M^\prime=\frac{2\pi}{\gamma}$,
$\tilde{N}=-5$ and $\tilde{\alpha}=-1$ while $\eta=\frac{1}{2}$.}
\end{figure}

Based on Eq.~(\ref{T}), if $\frac{\tilde{\alpha}}{\tilde{N}}>0$
(or equally $\tilde{\alpha}<0$), then $T=0$ for
$S=S_m\equiv(\frac{\tilde{\alpha}}{\tilde{N}})^\frac{2}{\eta}$, and
thus the system can obtain negative temperatures \cite{path}. In fact, this situation is very similar to a
system in which magnetic dipoles are located in the direction of
the external magnetic field $B$ \cite{path}. In Figs.~(\ref{fig4})
and~(\ref{fig5}), temperature, heat capacity and energy have been
plotted for $\eta=\frac{1}{2}$, leading to $S_0=16S_m$. In this manner, both the heat capacity and
temperature are negative while $0<S<S_m$. They will simultaneously
obtain their positive values for $S_m<S<S_0$. For $S>S_0$,
although temperature is positive, heat capacity is again negative, thus
signalling an unstable state \cite{path,25}.


Finally, it should again be noted that, unlike
Refs.~\cite{1,2,3,4,5,6,7,8,9,10,11,12,13,14,15,16,17,18,19,20,21,22,23,24,25},
we used the Misner-Sharp energy, in full agreement with the unified
first law of thermodynamics \cite{Moradpour:2016fur}, as the
thermodynamic potential in our calculations.
\section{Considerations}

We studied some thermodynamic properties of black holes in Rastall
gravity. The Misner-Sharp mass of Rastall black holes has been used in
our approach so to be compatible with the unified first law of
thermodynamics. Our investigation shows that the difference
between the Misner-Sharp mass of Rastall black holes and their
counterparts in Einstein's gravity will be decreased by reducing the size of
black hole. The behavior of the thermodynamic pressure of black holes
has also been studied showing that a non-minimal coupling between
geometry and matter fields in the Rastall way can lead to notable
effects on the pressure of the system. We finally investigated the
possibility of occurrence of phase transitions for Rastall black
holes. Like general relativity \cite{25}, a lower bound
for the horizon radius ($r_0$) was obtained, indicating that the
heat capacity of black holes with radius smaller than $r_0$ is
negative. This means that such black holes are unstable.


\section*{Acknowledgements}
IPL is supported by Conselho Nacional de Desenvolvimento
Cient\'ifico e Tecnol\'ogico (CNPq-Brazil) by the grant No.
150384/2017-3. The work of H. Moradpour has been supported
financially by Research Institute for Astronomy \& Astrophysics of
Maragha (RIAAM) under research project No. $1/5237-6$. JPMG is supported by CAPES (Coordena\c c\~ao de Aperfei\c coamento de Pessoal de N\'ivel Superior.


\section*{References}
\begin{thebibliography}{99}

\bibitem{Kiselev:2002dx}
  V.~V.~Kiselev,
  Class.\ Quant.\ Grav.\  {\bf 20}, 1187 (2003).

\bibitem{rastall} P. Rastall, Phys. Rev. D {\bf 6}, 3357 (1972).

\bibitem{Heydarzade:2017wxu}
  Y.~Heydarzade, F.~Darabi,
  Phys.\ Lett.\ B {\bf 771}, 365 (2017).

\bibitem{genras} H. Moradpour, Y. Heydarzade, F. Darabi, I. G. Salako, Eur. Phys. J. C {\bf 77}, 259 (2017).
\bibitem{PRL} T. Josset, A. Perez, Phys. Rev. Lett. 118, 021102 (2017).
\bibitem{REPJC} M. S. Ma, R. Zhao, Eur. Phys. J. C 77, 629 (2017)

\bibitem{deMello:2014hra}
  E.~R.~Bezerra de Mello, J.~C.~Fabris, B.~Hartmann,
  Class.\ Quant.\ Grav.\  {\bf 32}, no. 8, 085009 (2015).

\bibitem{Oliveira:2015lka}
  A.~M.~Oliveira, H.~E.~S.~Velten, J.~C.~Fabris and L.~Casarini,
  Phys.\ Rev.\ D {\bf 92}, no. 4, 044020 (2015).

\bibitem{Bronnikov:2016odv}
  K.~A.~Bronnikov, J.~C.~Fabris, O.~F.~Piattella and E.~C.~Santos,
  Gen.\ Rel.\ Grav.\  {\bf 48}, no. 12, 162 (2016).

\bibitem{vel} A. M. Oliveira, H. E. S. Velten, J. C. Fabris, L.
Casarini, Phys. Rev. D 93, 124020 (2016).

\bibitem{Heydarzade:2016zof}
  Y.~Heydarzade, H.~Moradpour and F.~Darabi,
  Can .Jour. Phys.

\bibitem{Licata} I. Licata, H. Moradpour, C. Corda, Int. Jour. Geo. Meth. Mod. Phys. Vol. 14, 1730003 (2017).
\bibitem{gaus} E. Spallucci, A. Smailagic, arXiv:1709.05795.

\bibitem{Capone:2009xm}
  M.~Capone, V.~F.~Cardone and M.~L.~Ruggiero,
  Nuovo Cim.\ B {\bf 125}, 1133 (2011).

\bibitem{Batista:2011nu}
  C.~E.~M.~Batista, M.~H.~Daouda, J.~C.~Fabris, O.~F.~Piattella and D.~C.~Rodrigues,
  Phys.\ Rev.\ D {\bf 85}, 084008 (2012).

\bibitem{Silva:2012gn}
  G.~F.~Silva, O.~F.~Piattella, J.~C.~Fabris, L.~Casarini and T.~O.~Barbosa,
  Grav.\ Cosmol.\  {\bf 19}, 156 (2013).

\bibitem{Santos:2014ewa}
  A.~F.~Santos and S.~C.~Ulhoa,
  Mod.\ Phys.\ Lett.\ A {\bf 30}, no. 09, 1550039 (2015).

\bibitem{hm} H. Moradpour, Phys. Lett. B {\bf757}, 187 (2016).

\bibitem{Yuan:2016pkz}
  F.~F.~Yuan and P.~Huang,
  Class.\ Quant.\ Grav.\  {\bf 34}, no. 7, 077001 (2017).

\bibitem{RHAG} Z. Haghani, T. Harko, S. Shahidi, arXiv:1707.00939.
\bibitem{p0} G. Q. Li, Phys. Lett. B 735, 256 (2014).
\bibitem{p1} B. Majeed, M. Jamil, P. Pradhan, AHEP, 2015, 124910 (2015).
\bibitem{p2} K. Ghaderi, B. Malakolkalami, Nuc. Phys. B 903, 10 (2016).
\bibitem{p3} K. Ghaderi, B. Malakolkalami, Astrophys. Space Sci. 361, 161 (2016).
\bibitem{p4} K. Ghaderi, B. Malakolkalami, Astrophys. Space Sci. 362, 163 (2017).
\bibitem{p5} Z. Xu, X. Hou, J. Wang, arXiv:1610.05454.
\bibitem{1} S. W. Hawking, D. N. Page,
  Commun.\ Math.\ Phys.\  {\bf 87}, 577 (1983).
\bibitem{2} J. Maldacena, Adv. Theor. Math. Phys. 2, 231 (1998).
\bibitem{3} E. Witten, Adv. Theor. Math. Phys. 2, 505 (1998).
\bibitem{4} A. Sahay, T. Sarkar, G. Sengupta, JHEP, 2010, 125 (2010).
\bibitem{5} R. Banerjee, S. K. Modak, S. Samanta, Eur. Phys. J. C, 70, 317 (2010).
\bibitem{6} R. Banerjee, S. K. Modak, S. Samanta, Phys. Rev. D 84, 064024 (2011).
\bibitem{7} Q. J. Cao, Y. X. Chen, K. N. Shao, Phys. Rev. D 83, 064015 (2011).
\bibitem{8} R. Banerjee, D. Roychowdhury, Phys. Rev. D 85, 044040 (2012).
\bibitem{9} R. Banerjee, S. Ghosh, D. Roychowdhury, Phys. Lett. B 696, 156 (2011).
\bibitem{10} R. Banerjee, D. Roychowdhury, JHEP, 2011, 4 (2011).
\bibitem{11} R. Banerjee, S. K. Modak, D. Roychowdhury, JHEP, 2012, 125 (2012).
\bibitem{12} S. W. Wei, Y. X. Liu, Eur. Phys. Lett, 99, 20004 (2012).
\bibitem{13} B. R. Majhi, D. Roychowdhury, Class. Quantum. Grav. 29, 245012 (2012).
\bibitem{14} W. Kim, Y. Kim, Phys. Lett. B 718, 687 (2012).
\bibitem{15} Y. D. Tsai, X. N. Wu, Y. Yang, Phys. Rev. D 85, 044005 (2012).
\bibitem{16} F. Capela, G. Nardini, Phys. Rev. D 86, 024030, (2012).
\bibitem{17} D. Kubiznak, R. B. Mann, JEHP, 2012, 33 (2012).
\bibitem{18} C. Niu, Y. Tian, X.-N. Wu, Phys. Rev. D 85, 024017 (2012).
\bibitem{19} A. Lala, D. Roychowdhury, Phys. Rev. D 86, 084027 (2012).
\bibitem{20} A. Lala, AHEP, 2013, 918490 (2013).
\bibitem{21} S. W. Wei, Y. X. Liu, Phys. Rev. D 87, 044014 (2013).
\bibitem{22} M. Eune, W. Kim, S. H. Yi, JHEP, 2013, 20 (2013).
\bibitem{23} M. B. J. Poshteh, B. Mirza, Z. Sherkatghanad, Phys. Rev. D 88, 024005 (2013).
\bibitem{24} J. X. Mo, W. B. Liu, Phys. Lett. B 727, 3361 (2013).
\bibitem{25} J. X. Mo, W. B. Liu, AHEP. 2014, 739454 (2014).
\bibitem{Moradpour:2016fur}
  H.~Moradpour, I.~G.~Salako,
  Adv.\ High Energy Phys.\  {\bf 2016}, 3492796 (2016).
\bibitem{Arnowitt:1959ah}
  R.~L.~Arnowitt, S.~Deser, C.~W.~Misner,
  Phys.\ Rev.\  {\bf 116}, 1322 (1959).

\bibitem{Bondi:1962px}
  H.~Bondi, M.~G.~J.~van der Burg, A.~W.~K.~Metzner,
  Proc.\ Roy.\ Soc.\ Lond.\ A {\bf 269}, 21 (1962).

\bibitem{Sachs:1962wk}
  R.~K.~Sachs,
  Proc.\ Roy.\ Soc.\ Lond.\ A {\bf 270}, 103 (1962).

\bibitem{Misner:1964je}
  C.~W.~Misner, D.~H.~Sharp,
  Phys.\ Rev.\  {\bf 136}, B571 (1964).

\bibitem{Vikman:2004dc}
  A.~Vikman,
  Phys.\ Rev.\ D {\bf 71}, 023515 (2005).

\bibitem{Sahni:2004ai}
  V.~Sahni,
  Lect.\ Notes Phys.\  {\bf 653}, 141 (2004).

  \bibitem{Abbott:2016blz}
  B.~P.~Abbott {\it et al.} [LIGO Scientific and Virgo Collaborations],
  Phys.\ Rev.\ Lett.\  {\bf 116}, no. 6, 061102 (2016).

\bibitem{Abbott:2016nmj}
  B.~P.~Abbott {\it et al.} [LIGO Scientific and Virgo Collaborations],
  Phys.\ Rev.\ Lett.\  {\bf 116}, no. 24, 241103 (2016).

\bibitem{Abbott:2017vtc}
  B. P. Abbott {\it et al.} [LIGO Scientific and VIRGO Collaborations],
  Phys.\ Rev.\ Lett.\  {\bf 118}, no. 22, 221101 (2017).

\bibitem{Abbott:2017oio}
  B.~P.~Abbott {\it et al.} [LIGO Scientific and Virgo Collaborations],
  Phys.\ Rev.\ Lett.\  {\bf 119}, no. 14, 141101 (2017).

\bibitem{Perlmutter:1998np}
  S. Perlmutter {\it et al.} [Supernova Cosmology Project Collaboration],
  Astrophys.\ J.\  {\bf 517}, 565 (1999).

\bibitem{Caldwell:1999ew}
  R. R. Caldwell,
  Phys.\ Lett.\ B {\bf 545}, 23 (2002).

\bibitem{path} R. K. Pathria, P. D. Beale, \textit{Statistical Mechanics (Third
Edition)} (30 Corporate Drive, Suite 400, Burlington, MA 01803,
USA 2011).

\bibitem{revm} L. B. Szabados, Living Rev. Rel. 7, 4 (2004).
\bibitem{j1} M. Akbar, R. G. Cai, Phys. Lett. B {\bf648}, 243 (2007)
\bibitem{j2} M. Akbar, R. G. Cai, Phys. Lett. B {\bf635}, 7 (2006).
\bibitem{j3} M. Akbar, R. G. Cai, Phys. Rev. D {\bf75}, 084003 (2007).
\bibitem{j4} R. G. Cai, L. M. Cao, Phys. Rev. D {\bf75}, 064008 (2007).
\bibitem{j5} R. G. Cai, L. M. Cao, Nucl. Phys. B {\bf785}, 135 (2007).
\bibitem{j6} A. Sheykhi, B. Wang, R. G. Cai, Nucl. Phys. B {\bf779}, 1 (2007).
\bibitem{j7} A. Sheykhi, B. Wang, R. G. Cai, Phys. Rev. D {\bf76}, 023515 (2007).
\bibitem{j8} A. Sheykhi, J. Cosmol. Astropart. Phys. {\bf05}, 019 (2009).
\bibitem{j9} A. Sheykhi, Eur. Phys. J. C {\bf69}, 265 (2010).
\bibitem{j10} A. Sheykhi, Class. Quantum. Gravit. {\bf27}, 025007 (2010).
\bibitem{j11} R. G. Cai, N. Ohta, Phys. Rev. D {\bf81}, 084061 (2010).
\bibitem{j12} A. Sheykhi, Phys. Rev. D {\bf87}, 024022 (2013).
\bibitem{j13} A. Sheykhi, M. H. Dehghani, R. Dehghani, Gen. Relativ. Gravit. {\bf46}, 1679 (2014).
\bibitem{ref2} R. G. Cai, L. M. Cao, Y. P. Hu, N. Ohta, Phys. Rev. D \textbf{80}, 104016 (2009).
\end{thebibliography}
\end{document}